\documentclass[sigconf]{acmart}

\usepackage{booktabs} 

\copyrightyear{2019}
\acmYear{2019} 
\setcopyright{iw3c2w3}
\acmConference[WWW '19]{Proceedings of the 2019 World Wide Web Conference}{May 13--17, 2019}{San Francisco, CA, USA}
\acmBooktitle{Proceedings of the 2019 World Wide Web Conference (WWW '19), May 13--17, 2019, San Francisco, CA, USA}
\acmPrice{}
\acmDOI{10.1145/3308558.3314119}
\acmISBN{978-1-4503-6674-8/19/05}

\usepackage{balance}

\begin{document}

\title{XFake: Explainable Fake News Detector with Visualizations}

\author{Fan Yang$^{*1}$, Shiva K. Pentyala$^{*1}$, Sina Mohseni$^{*1}$, Mengnan Du$^{*1}$, Hao Yuan$^{*2}$, Rhema Linder$^{1}$, \and Eric D. Ragan$^{3}$, Shuiwang Ji$^{1}$, Xia (Ben) Hu$^{1}$}\thanks{$*$ Those authors contributed equally in developing the system.}

\affiliation{
\institution{$^1$Department of Computer Science and Engineering, Texas A\&M University}
\institution{$^2$School of Electrical Engineering and Computer Science, Washington State University}
\institution{$^3$Department of Computer \& Information Science \& Engineering, University of Florida}
}
\email{{nacoyang, pk123, sina.mohseni, dumengnan, rhema, sji, xiahu}@tamu.edu, hao.yuan@wsu.edu, eragan@ufl.edu}

\renewcommand{\shortauthors}{F. Yang et al.}

\begin{abstract}
In this demo paper, we present the XFake system, an explainable fake news detector that assists end-users to identify news credibility. To effectively detect and interpret the fakeness of news items, we jointly consider both attributes (e.g., speaker) and statements. Specifically, \textit{MIMIC}, \textit{ATTN} and \textit{PERT} frameworks are designed, where MIMIC is built for attribute analysis, ATTN is for statement semantic analysis and PERT is for statement linguistic analysis. Beyond the explanations extracted from the designed frameworks, relevant supporting examples as well as visualization are further provided to facilitate the interpretation. Our implemented system is demonstrated on a real-world dataset crawled from \textit{PolitiFact}\footnote{https://www.politifact.com/}, where thousands of verified political news have been collected. 
\end{abstract}



\maketitle

\section{Introduction}

With the prevalence of web applications, it has become much easier for the public to get access to various news items through different channels, such as news websites and social media. However, this kind of convenience also brings about the wide spread of fake news at the same time~\cite{zhou2018fake}. 
Fake news is detrimental to both individuals and society, and it is usually generated as hoaxes to mislead public through false or biased information.  
In 2016, the term "Fake News" was even shown as the word of the year by Macquarie Dictionary. 

\begin{figure}[tbp] 
\centering\includegraphics[width=3.35in]{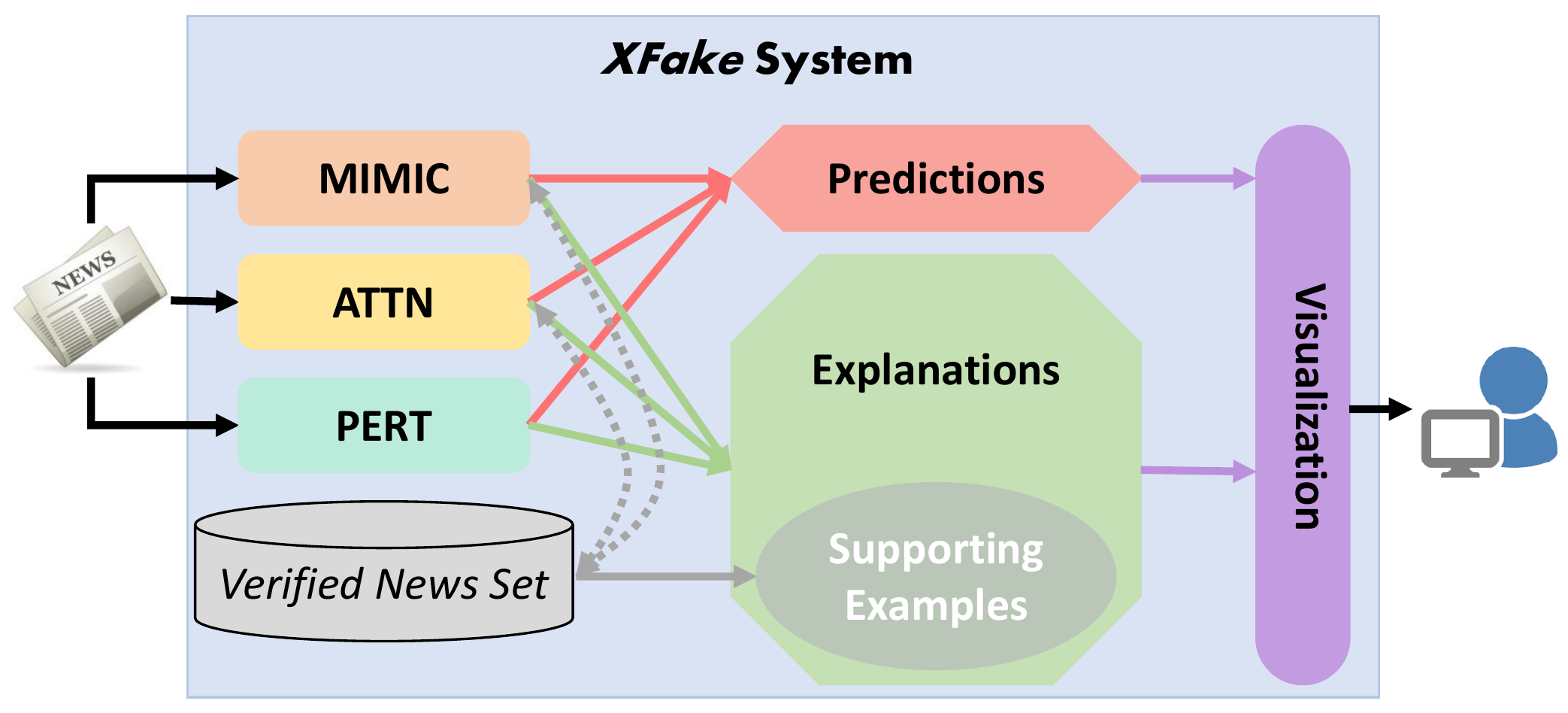} 
\vspace{-0.6cm}
\caption{The architecture of XFake system.} 
\vspace{-0.5cm}
\end{figure}

Though increasingly relevant, effective fake news detection is still considered to be challenging due to the following two aspects. First, fakeness of news could come from diversified perspectives~\cite{rubin2015deception}, which is beyond the boundaries of traditional textual analysis. For example, a fake news item may result from its contexts or speaker, and it may not directly relate to its contents. The source of fakeness is highly dependent on each specific instance. Second, fakeness detection results need further explanation, which is important and necessary for users' final decisions~\cite{gunning2017explainable}. The explanation could provide evidences on model predictions and further help users better understand why certain news items are classified as fake. 

Considering these two challenges, we design and implement an explainable system, named \textbf{XFake}, on fake news detection task. Particularly, three different frameworks (i.e., \textit{MIMIC}, \textit{ATTN} and \textit{PERT}) are designed to analyze news items from different perspectives. With MIMIC, we aim to judge each news item from its attribute information, which may include news context, speaker and etc. For ATTN, our goal is to investigate each news statement through its semantic meaning. By employing PERT, we aim to study the linguistic features of news statements for detection. All three designed frameworks are self-explainable, where relevant explanations could be extracted for interpreting detection results. Besides, we also provide supporting examples for each news instance by corresponding matching algorithm, and visualize detection results for user-friendly interaction. Both supporting examples and visualizations further assist users to understand why the system make a certain prediction on a given news. Overall, XFake system is capable of not only giving prediction scores on news fakeness, but also providing relevant explanations as prediction evidences. With the aid of XFake, users would be convenient to identify the news credibility. 

To best of our knowledge, XFake is the first explainable fake news detector over multiple perspectives. Corresponding system architecture and demonstration details will be respectively introduced in the following section 2 and section 3.  

\section{XFAKE Architecture}

The architecture of XFake is illustrated in Figure 1. The system input is the targeting news with attributes, and the output contains both prediction and explanation results. The whole system is trained on a verified news set. All components will be introduced as below. 

\vspace{1ex}\noindent
\textbf{MIMIC Framework.} MIMIC is designed to analyze the news attributes. Enlightened by~\cite{wang2017liar}, we employ a deep neural network as the teacher, and use a shallow model as the student to emulate the teacher's performance for better explainability. Basically, the overall idea is to mimic the performance of neural networks with tree ensemble models, so that we can keep the good performance from neural networks and good explainability from tree ensemble models simultaneously. The structure of MIMIC is illustrated in Figure 2. Specifically, we use GloVe~\cite{pennington2014glove} as our word embedding scheme in MIMIC. After obtaining the soft labels from the teacher model, we further use them to train a student model consisting of 80 decision trees. By calculating the node importance of each attribute in the ensemble trees, we could get the significance of each attribute in the input news instance. Further, through the constructed trees and importance scores, users could know what happened inside, which provides an overall transparency towards the model itself. 
\begin{figure}[tbp] 
\centering\includegraphics[width=3.1in]{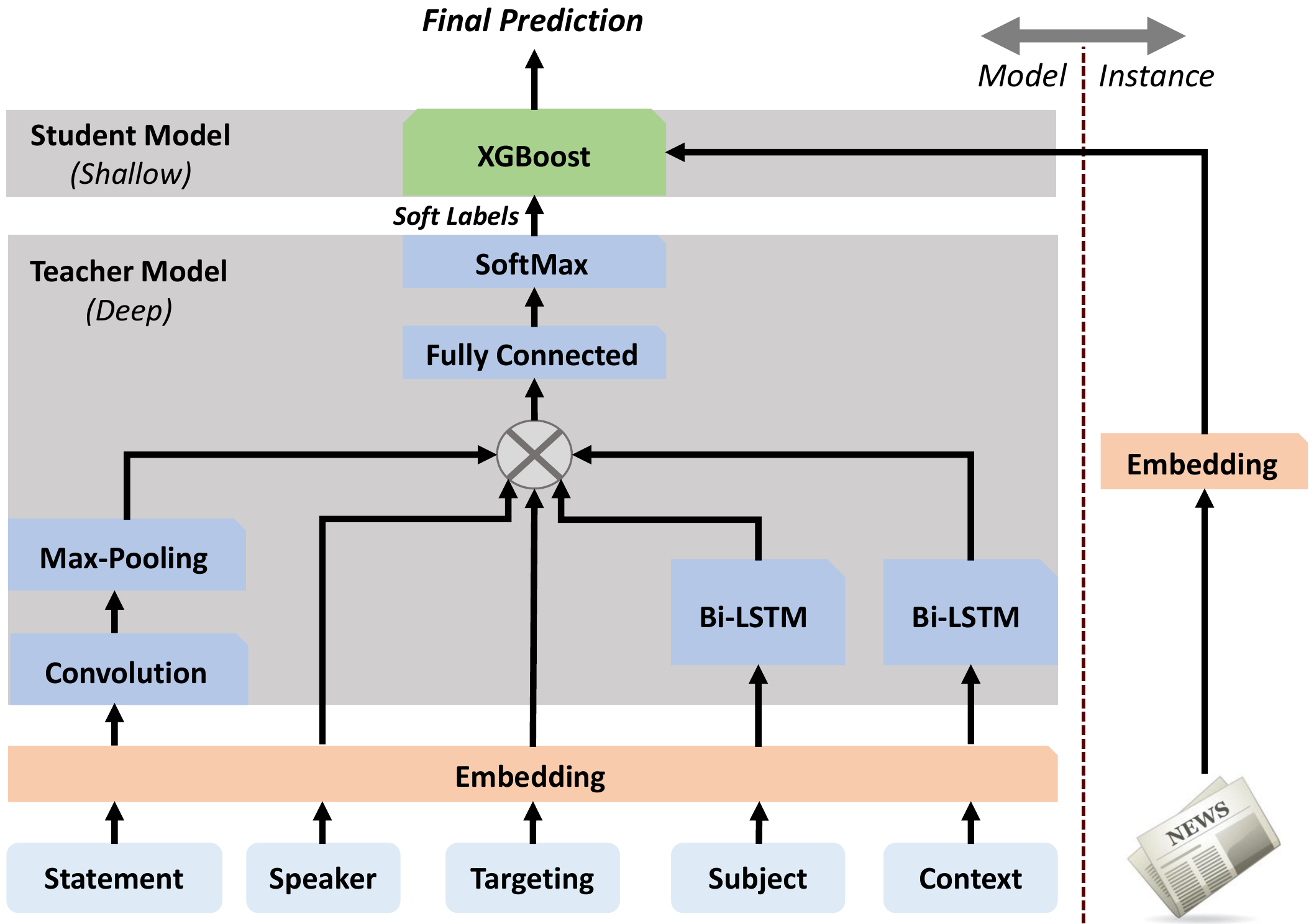} 
\vspace{-0.2cm}
\caption{The structure of MIMIC framework.} 
\vspace{-0.46cm}
\end{figure}

\vspace{1ex}\noindent
\textbf{ATTN Framework.} ATTN is designed to analyze news statement simply from semantic perspective. To build an explainable model for semantic analysis with good performance,  we employ several techniques, including pre-trained word embedding, convolutional neural network~\cite{krizhevsky2012imagenet}, and self-attention mechanism~\cite{vaswani2017attention}. Self-attention is used because it can capture global relationships between different words efficiently. In addition, the weight matrix generated in attention mechanism is input dependent, which helps provide instance-level explanation. By applying different kernel sizes for the convolutional network, we can get explanations based on one-gram, two-gram or three-gram analysis. The overall illustration of ATTN is given by Figure 3. In ATTN, we employ word2vec~\cite{mikolov2013distributed} as the pre-trained embeddings and each vector representation has a dimension of 300 (i.e. $E=300$). Each spatial location learns a 512-dimensional vector representation for each word (i.e. $D=512$). 

\vspace{1ex}\noindent
\textbf{PERT Framework.} PERT is designed for news statement analysis as well, but it is uniquely from a linguistic perspective. For effective analysis, we employ eight linguistic features, including \textit{Adjective ratio}, \textit{Noun ratio}, \textit{Verb ratio}, \textit{Propn} (such as Google, etc.) \textit{ratio}, \textit{Sentiment score}, \textit{Normalized text length}, \textit{Whether contains the mark "?"}, \textit{Whether contains the mark "!"}~\cite{herzallah2018feature}. For each news item in the training set, we extract its linguistic features and train an XGBoost~\cite{chen2016xgboost} classifier using these features. The trained XGBoost is then used to make predictions for new items. Further, we use perturbation-based method to provide explanations. The idea is that feature importance can be measured by observing how much the score (e.g., accuracy) decreases when the feature is not available. To this end, we can remove a feature from the dataset, and then re-train the classifier for score checking. Since re-training may be computationally expensive, we replace the feature value with random noise, drawn from the same distribution as the original one, instead of removing. The computed prediction difference is utilized as the significance score for the corresponding feature. 
\begin{figure}[tbp] 
\centering\includegraphics[width=2.8in]{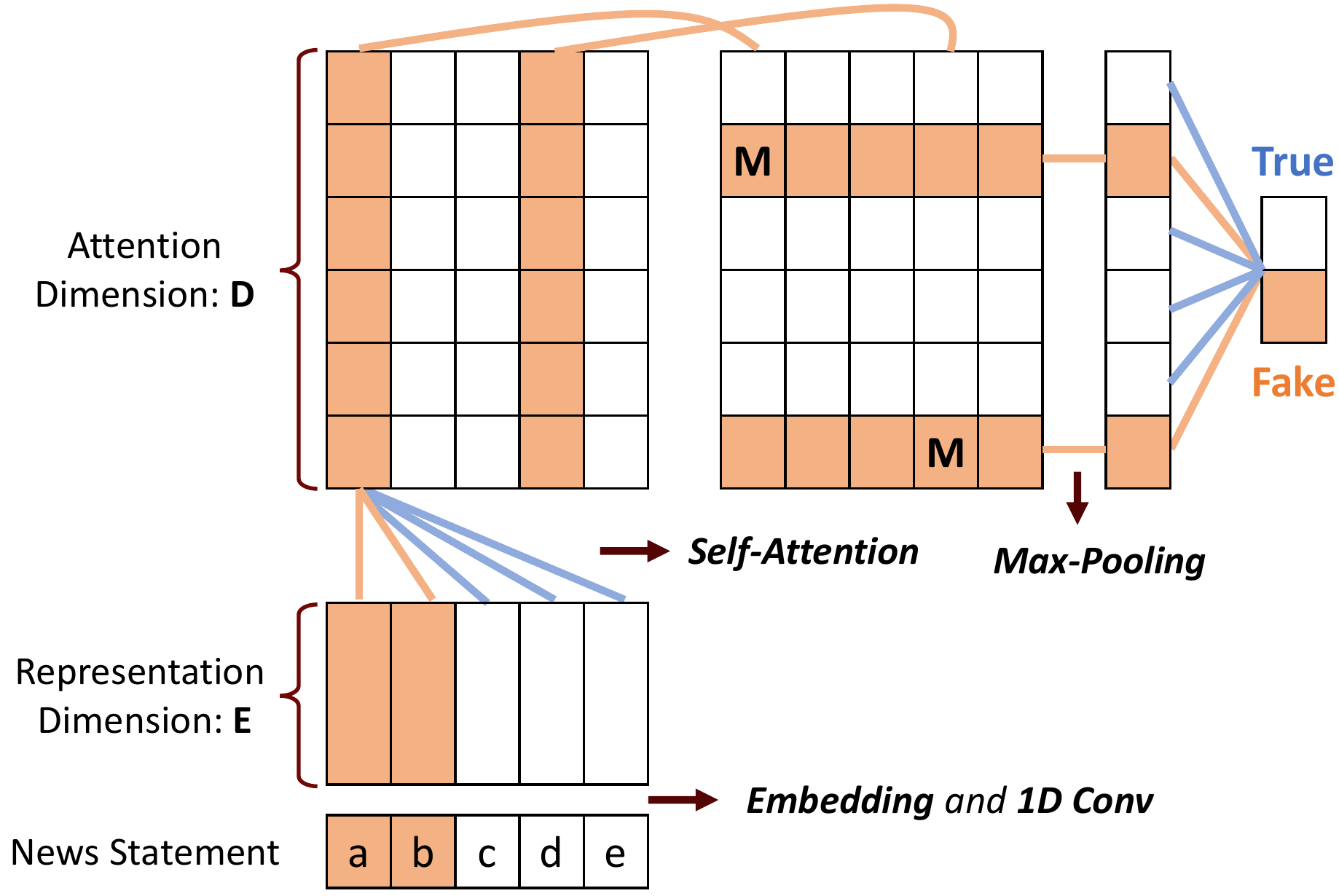} 
\vspace{-0.2cm}
\caption{The illustration of ATTN framework.} 
\vspace{-0.2cm}
\end{figure} 
\begin{figure}[tbp] 
\centering\includegraphics[width=3.0in]{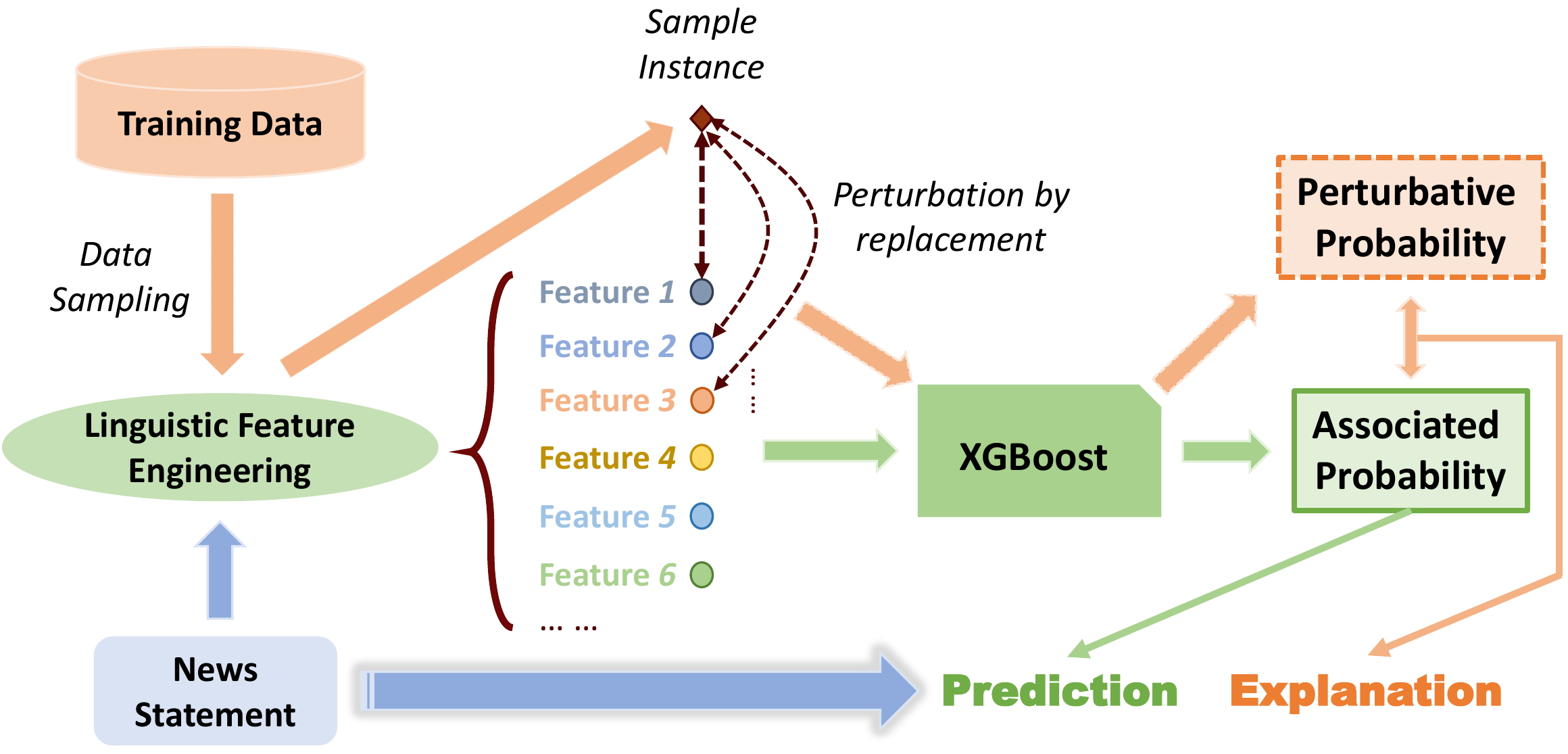} 
\vspace{-0.3cm}
\caption{The mechanism of PERT framework.} 
\vspace{-0.4cm}
\end{figure}

\vspace{1ex}\noindent
\textbf{Prediction \& Explanation.} XFake outputs the prediction of news as a unified score, indicating its probability to be fake. The prediction result is the ensemble output from MIMIC, ATTN and PERT, where different results are combined together in a weighted sum manner. The weights for combination are tuned on a validation set, according to the performance of different frameworks. The higher the prediction score, the higher the probability for fake news. As for the explanation from XFake, it is mainly extracted from three different perspectives enabled by MIMIC, ATTN and PERT. MIMIC could provide explanation by key components of news, ATTN is able to explain predictions by word/phrase attribution, and PERT generates explanation by linguistic features. Beyond this, XFake also explains predictions from data perspective with supporting examples. These examples are generated from both MIMIC and ATTN by retrieving training samples given corresponding explanations (i.e. key attributes or important words/phrases). Showing these training samples which are highly similar to the input news would be helpful for users to understand the working patterns of XFake. Besides, we also assign different similarity scores for different supporting examples based on the matching extent between the input and support. Supporting examples with higher scores would be considered more informative in delivering explanation. 

\vspace{1ex}\noindent
\textbf{Visualization.} To further facilitate the explanability of XFake, we visualize the outputs of both prediction and explanation by \textit{D3 JavaScript}. Visualization mainly lies in three aspects. First, for numerical values, such as prediction score and attribute significance, we visualize them by histograms, which straightforwardly indicate the results and influences. Second, to enhance the explanability for word/phrase attribution, we visualize the outputs by highlighting important words/phrases with heatmaps, where the darkness positively relates to the importance of word/phrase. Third, for better model explanability, the ensemble trees are visualized with interactive diagrams which are capable of showing both overall structure and specific activated paths (depending on the input). Through those visualization schemes, users could have a better sense towards XFake about why certain news are classified as fake or true. 

\section{System Demonstration}

XFake is implemented by \textit{Python} and deployed in \textit{FLASK} with an \textit{HTML} front end. To demonstrate the system, we will first introduce the data source and then go through a specific use case of XFake. 

\subsection{Data Source}

The news data, which is used to train, tune and evaluate XFake, comes from a political fact-checking website, named PolitiFact. It is a Pulitzer prize-winning website containing tons of political news with diversified categories. The reasons why we employ this data source are in three folds. First, PolitiFact provides professional justification and fine-grained labels for all news items, where the core principles in independence, transparency and fairness guarantee its high credibility among the public. Second, the news collected by PolitiFact have various attribute information, which directly meets our data requirement for analysis. Third, raw data in PolitiFact can be effectively crawled through an available API~\footnote{http://static.politifact.com.s3.amazonaws.com/api/v2apidoc.html}, and it is convenient to obtain the customized dataset for system implementation. 

When processing the data, we only keep the attributes which are highly related to news fakeness. Specifically, the maintained attributes for XFake include \textit{Subject}, \textit{Context}, \textit{Speaker}, \textit{Targeting} and \textit{Statement}, although some news may not have all five attributes. Besides, to effectively measure the news fakeness and train the system, we transform the original multi-class data to binary data where each news item is labelled as either \textit{True} or \textit{False}\footnote{A news item labelled as False is regarded as the fake news.}. Particularly, labels with \textit{Mostly True}, \textit{Half True}, \textit{No Flip}, \textit{Half Flip} are switched to label \textit{True}, and labels with  \textit{Mostly False}, \textit{Pants On Fire}, \textit{Full Flop} are switched to label \textit{False}.  Furthermore, as for data statistics, there are \textit{5104} news crawled in total, which are evenly distributed. The training, validation, test set of XFake respectively contain \textit{4083}, \textit{510}, \textit{511} items. The validation set is used to test the performance of different frameworks and further tune the hyper-parameters, while the test set is employed to conduct the overall evaluation. 

\subsection{Use Cases}

\begin{figure*}[tbp]
\centering\includegraphics[width=0.99\textwidth]{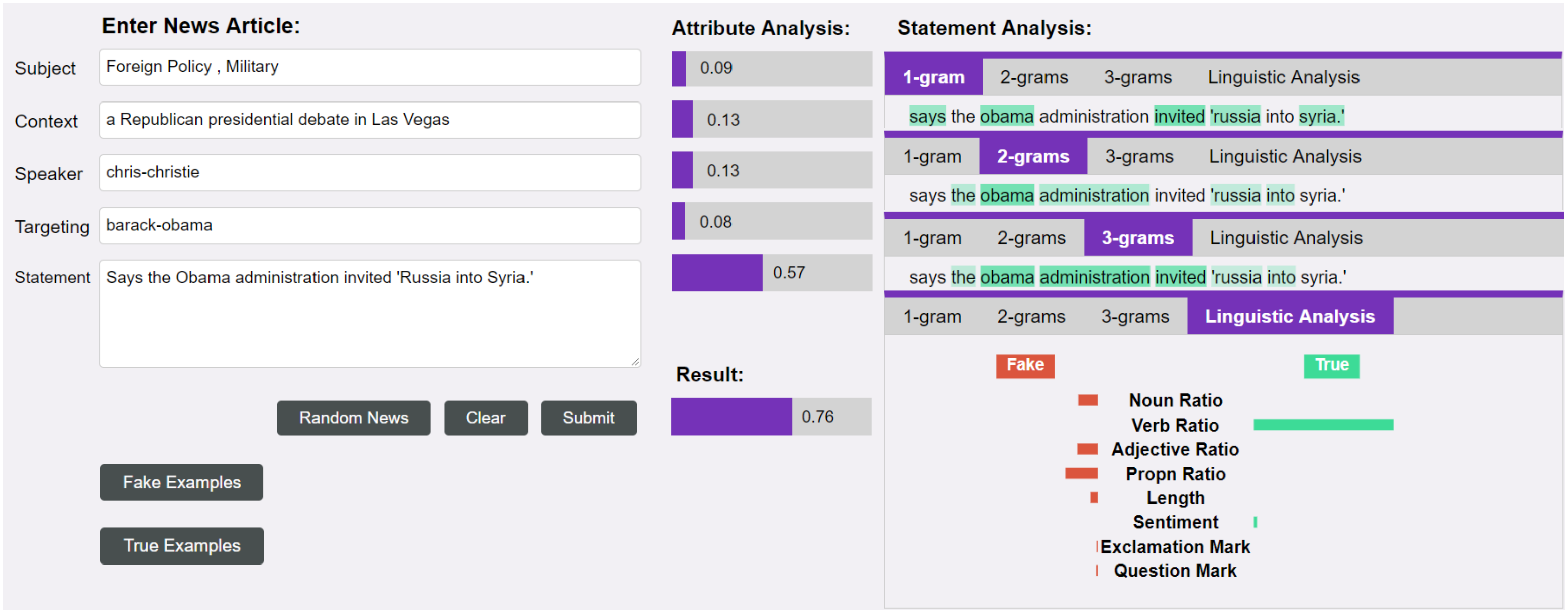} 
\vspace{-0.29cm}
\caption{Prediction and explanation from XFake.} 
\end{figure*}
\begin{figure*}[tbp]
\centering\includegraphics[width=0.99\textwidth]{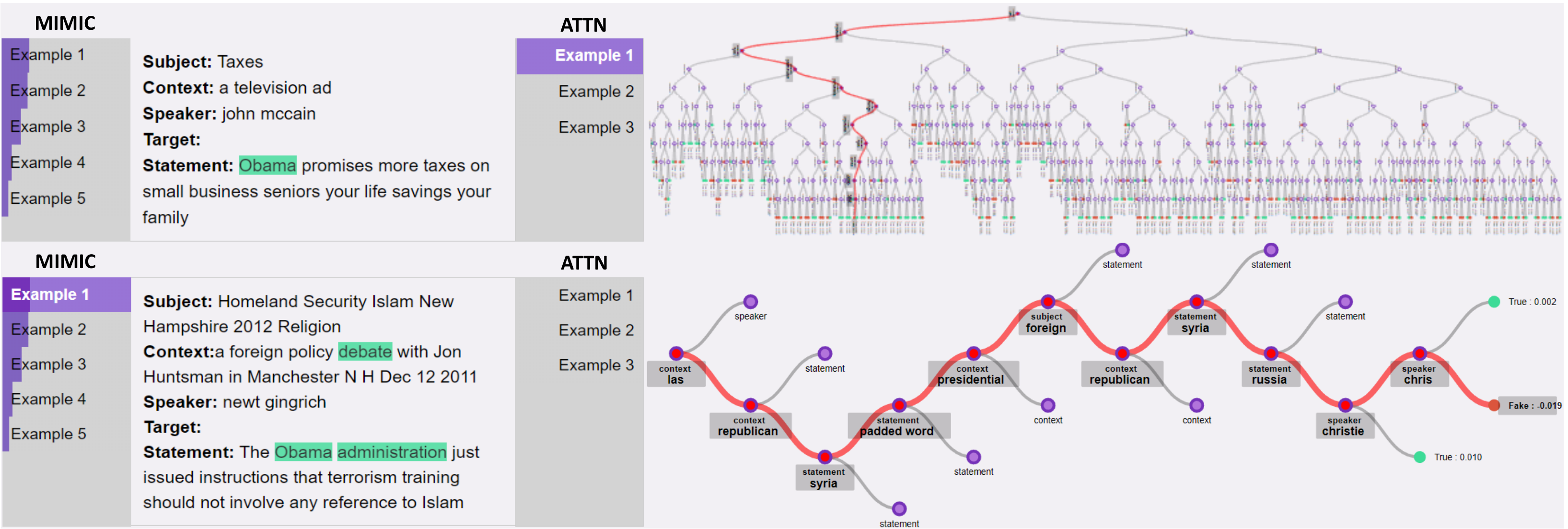} 
\vspace{-0.29cm}
\caption{Supporting examples and ensemble trees from XFake.} 
\end{figure*}

With the testing on validation set, we obtain \textit{67.1\%}, \textit{67.3\%}, \textit{53.2\%} accuracy respectively for MIMIC, ATTN and PERT framework. By normalizing weights according to the performance, we fix the coefficients \textit{0.36}, \textit{0.36}, \textit{0.28} in weighted sum correspondingly for all use cases. Considering the fake news identification scenario, we show a specific case demonstration of XFake as follows. 

Illustrated by Figure 5, users start by inputting news into the text boxes. In XFake, we provide a button "Random News" to help users explore the system, which is used to retrieve random items from our test set. Similarly, buttons "Fake Examples" and "True Examples" are also provided to help quickly access some representative fake and true news. After clicking "Submit", users would obtain all the outputs including both prediction and explanation in a few seconds. As for the prediction of the example in Figure 5, we get the score 0.76, which means that the given news has the probability 76\% to be fake. Regarding to the explanation, we can obtain it from both attribute and statement analysis. Aided by MIMIC, for this example, we know that "Statement" plays the most important role compared with others. Through ATTN, we can easily check those highlighted words, such as "invited" and "Russia", with different darkness, which would also show the contribution scores when mouse is hovering around. PERT gives users a clear view about which linguistic features contribute to fake and which to true. In the example, we observe that features "Propn Ratio", "Adjective Ratio" and "Noun Ratio" mainly contribute this news to be fake. 

XFake further provides supporting examples and visualized trees for users to better understand the system. As shown in Figure 6, we give two supporting news for instance, where one is retrieved based on the important attributes (Context \& Statement) from MIMIC and the other is obtained by matching significant word ("Obama") from ATTN. For the support extraction with MIMIC, we also attach a similarity score, indicating how much attribute information it overlaps with the input one. Besides, 80 decision trees are visualized with interactive diagrams and highlight the activated path of each tree regarding to the input. In Figure 6, we only show one decision tree for example. We can see that each decision tree can be expanded or compressed flexibly, which allows users to track the decision process closely. Given a certain news, each decision tree has only one activated path, corresponding to one specific decision attached at the end of the path with relevant contribution score. Those visualized trees largely enhance the model explainability of XFake. 

To demonstrate the effectiveness of XFake in real cases, we conduct relevant human evaluations by Amazon Mechanical Turk (AMT), with \textit{147} valid testing users in total covering diversified gender, age and education level. The involved user tasks include \textit{Fact Check} and \textit{Prediction Guess}, where the first one is to test the usefulness of the generated explanation and the second one is to indicate the users' understanding towards the system. The evaluation metrics are accuracy and time for user prediction. 
The human study shows a clear trade-off between the speed and accuracy regarding to generated explanations. On one hand, explanation does help users better understand and predict system behavior. On the other hand, explanation would take users more time to review and interpret detection results for benefits. Due to space constraints, we limit the details of the human evaluation study in this paper. 

\section{Conclusions and Future Work}

We design and implement an explainable fake news detector, named XFake, to help end-users identify the news credibility. XFake is capable of jointly analyzing news items from multiple perspectives, and can effectively provide explanation through attribute importance, word/phrase significance, linguistic features as well as relevant supporting examples. Corresponding visualizations further enhance the explanability of XFake and facilitate the decision making for users. In future, applying XFake to multi-source (i.e., not limited to the PolitiFact website) or multi-modal (i.e., not limited to textual form) news data could be a promising extension for the system. 

\begin{acks}
The authors thank the anonymous reviewers for their helpful comments and the funding agencies for their generous supports.
The work is in part supported by DARPA grant N66001-17-2-4031. The views and conclusions in this paper are those of
the authors and should not be interpreted as representing any funding agencies.
\end{acks}

\bibliographystyle{ACM-Reference-Format}
\balance
\bibliography{WebConf19_XFake_CR}

\end{document}